\documentclass{jetpl}
\usepackage[cp1251]{inputenc}
\usepackage{bm}
\usepackage{amsmath}
\usepackage{amssymb}
\usepackage{amscd}
\usepackage[dvips]{graphicx}
\usepackage{epsfig}

\twocolumn

\rus

\title{Shot noise measurements in a wide-channel transistor near pinch-off}

\rtitle{\ldots}

\sodtitle{}

\author{V.S.~Khrapai, D.V.~Shovkun}

\rauthor{}

\sodauthor{}

\address{Institute of Solid State Physics, Russian Acaddemy of Sciences, 142432 Chernogolovka, Russian Federation}
\dates{}{*}

\abstract{We study a shot noise of a wide channel gated
high-frequency transistor at temperature of 4.2K near pinch-off.
In this regime, a transition from the metallic to the insulating
state is expected to occur, accompanied by the increase of the
partition noise. The dependence of the noise spectral density on
current is found to be slightly nonlinear. At low currents, the
differential Fano factor is enhanced compared to the universal
value 1/3 for metallic diffusive conductors. We explain this
result by the effect of thermal fluctuations in a nonlinear regime
near pinch-off, without calling for the enhanced partition noise.}


\begin{document}

\maketitle

A current $I$ flowing in a two-terminal conductor placed in
external electric circuit exhibits fluctuations around it's mean
value $\overline{I}$. The second moment of the fluctuations is
related to the noise spectral density
$\left.\overline{(I-\overline{I})^2}\right|_{\Delta f}\equiv
S_I\Delta f$, where $\Delta f$ is a measurement bandwidth. In the
absence of current ($\overline{I}=0$) the noise is related to
thermal fluctuations of the occupation number of the electronic
states, known Johnson-Nyquist noise (JN-noise). In this case
$S_I=4k_BTR^{-1}$, where $k_B,T$ and $R$ are, respectively, the
Boltzman constant, the temperature and the resistance of the
conductor. Away from the equilibrium, when the voltage drop across
the conductor is high enough $|eV|\gg k_BT$, and in the absence of
dissipation inside the conductor and spurious noises, the current
fluctuations are caused by the discreteness of the elementary
charge $e$~\cite{schottky}. This noise is referred to as shot
noise and for a voltage-biased conductor has a spectral density of
$S_I=2F|e\overline{I}|$, where $F$ is called a Fano-factor.

In a non-interacting system in the linear transport regime, the
shot noise is caused by a partition of incident carriers, which
can be viewed as quantum effect~\cite{blanterreview}. In this
case, the Fano-factor is determined by the distribution of the
eigen-channel transparencies ($T_n$) of the conductor ${F=
\overline{T_n(1-T_n)}/\overline{T_n}}\leq1$~\cite{blanterreview}.
The noise is strongest ($F=1$) in the Poissonian regime, which is
obtained when all $T_n\ll1$. In a quasi one dimensional metallic
diffusive conductor the distribution of $T_n$ is
universal~\cite{beenaker} and $F=1/3$, which has been
experimentally confirmed~\cite{henny}. The universality of the
value $F=1/3$ in metallic conductors has been proven to be
independent of geometry~\cite{nazarov}. Near the transition from
the metallic to the insulating state one expects an increase of
the partition noise to the Poissonian value~\cite{blanterreview},
although this regime haven't been studied experimentally.

We study the shot noise in a gated wide channel transistor near
pinch-off, where the transition to the insulating state is
expected to occur. The dependence of $S_I$ on current is slightly
non-linear in the shot noise regime. The differential Fano-factor
$F_D=(2|e|)^{-1}|dS_I/dI|$ is enhanced above the universal
metallic value $1/3<F_D\leq0.5$ for low $|I|$ and is close to this
value $F_D\approx1/3$ for higher $|I|$. We find that the
enhancement of $F_D$ is not necessarily related to the quantum
partition noise. In contrast, it can be explained by a classical
effect of thermal fluctuations in a strongly nonlinear transport
regime near pinch-off.

\begin{figure}
\begin{center}
\includegraphics[width=0.9\columnwidth]{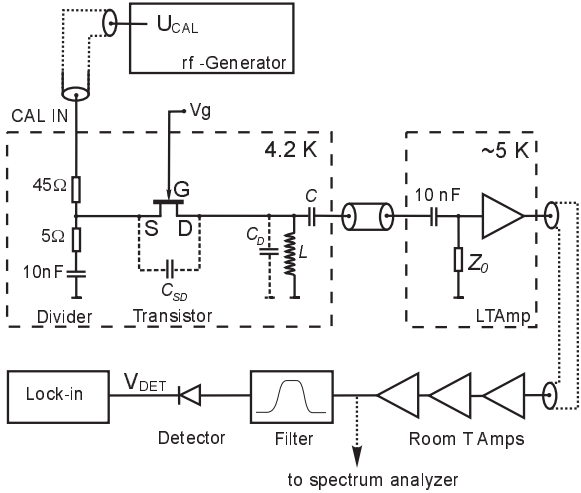}
\end{center}
\caption{Fig.~\ref{fig1}: The sketch of the setup. The low
temperature parts are shown by dashed boxes. The parameters of the
resonant $LC$ circuit and the undesired stray capacitances
$C_{SD}$ and $C_D$ (shown by dashed lines inside the 4.2K box) are
given in the text. The signal from the room temperature amplifiers
(Room T Amps) is either sent to the filter and detector for a
Lock-in measurement or to the spectrum analyzer (dashed line with
an arrow) for wide range spectra acquisition.}\label{fig1}
\end{figure}

The skecth of the measurement is shown in fig.~\ref{fig1}. The
sample and the cryogenic amplifier (LTAmp) are placed in a $^4$He
gas chamber with the walls maintained at 4.2K. The sample is below
the LTAmp and is connected to it with a 20~cm cable. A heat sink
connects the LTAmp to a liquid $^4$He bath. The actual temperature
of the LTAmp ($T_{LTA}$) measured with a thermometer is about
5.3K. The actual temperature of the sample is taken to be 4.2K,
consistent with the JN-noise measurement (see below). At the input
of the setup (CAL IN) a 50~$\Omega$ cable is connected to the
transistor source (S) via a divider. This input is used both for
driving a current $I$ and external rf calibration. A transistor
drain (D) is followed by an $L-C$ resonator, which is connected to
a  cryogenic amplifier (LTAmp) ($\approx20$dB gain). The resonator
serves to match a high-impedance of the sample and a low input
impedance $Z_0=50~\Omega$ of the LTAmp. The output of the LTAmp is
connected via a second 50~$\Omega$ cable to the input stage of the
room-temperature low-noise amplifiers (total gain of
3$\times$20~dB). Finally, the amplified signal is filtered with a
$\sim$30~MHz bandpass filter at the resonance frequency
$f_0\approx125$~MHz and rectified by a detector (8473C by Agilent
Technologies). The ac modulation of the rectified signal thanks to
a current modulation, gate voltage chopping or amplitude
modulation of the external rf is measured with the lock-in.
Alternatively, a spectrum analyzer can be used to analyze the
frequency spectra at the output of the room-temperature
amplifiers.
We study shot noise of a commercial pseudomorphic
AlGaAs/InGaAs/GaAs pHEMT ATF-35143 by Agilent with a gate length
(width) of 0.5~(400)~$\rm \mu m$. This transistor is known for
low noise at room temperature and is used as an active element in
cryogenic LTAmp's by us and other
authors~\cite{RoschierCryogenics}. We preformed the resistance
measurements in two such transistors (samples 1 and 2), and noise
measurements and calibration only in sample 2.
\begin{figure}
\begin{center}
\includegraphics[width=0.7\columnwidth]{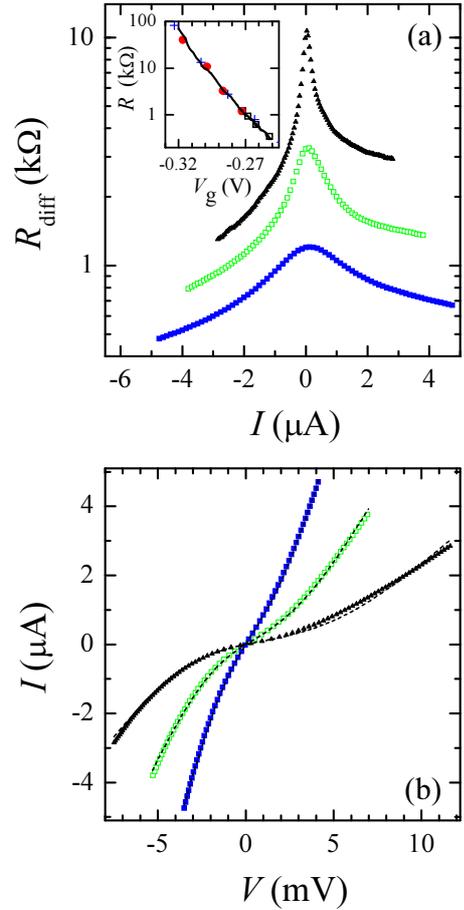}
\end{center}
\caption{Fig.~\ref{fig2}. (a) Differential resistance $R_{diff}$
as a function of current for three values of the gate voltage in
sample 2. Inset -- linear response S-D resistance as a function of
gate voltage for sample 1 (solid line) and three different states
of sample 2 (symbols). The $V_g$ axis for different sample states
was compensated for random shifts on the order of 50~mV. (b)
Experimental $I$-$V$ curves calculated from the data of (a) shown
by the same symbols as the corresponding data for $R_{diff}$ in
(a). Model fits (see text) are shown by dashed lines.}\label{fig2}
\end{figure}

Negative gate voltage $V_g<0$ is used to deplete the channel which
results in increase of the linear-response S-D resistance $R\equiv
dV/dI|_{I\rightarrow0}$. The huge aspect ratio of the gate
electrode allows to work near the pinch-off with high channel
resistivity in the range of ~M$\Omega/\Box$ while keeping a
reasonable $R$. The dependence $R(V_g)$ at 4.2~K is shown in the
inset of fig.~\ref{fig2} for sample 1 (solid line) and for several
states of sample 2 (symbols). In all cases, the behavior $R(V_g)$
is roughly exponential and is reproducible up to insignificant
random threshold voltage shifts (see caption). Hence, most likely,
the current is homogeneously distributed across the transistor
channel near pinch-off. Such a strong dependence $R(V_g)$ might
indicate that we enter the insulating phase near the
pinch-off~\cite{ando}. The measured differential resistance
$R_{diff}\equiv dV/dI$ is shown as a function of current in
fig.~\ref{fig2}a for three values of the gate voltage in sample 2.
These data are taken simultaneously with the shot noise
measurements presented below. $R_{diff}$ is maximum in the linear
response ($I=0$) and falls down at finite current. The reduction
is most pronounced at low $|I|$ and for more depleted channel.
$R_{diff}$ is an asymmetric function of current, which is related
to the capacitive population/depopulation of the channel at
negative/positive bias. The nonlinear $I$-$V$ curves numerically
calculated from these data are plotted in fig.~\ref{fig2}b. The
nonlinearity is somewhat similar to the behavior of the $I$-$V$
curves in the insulator breakdown regime~\cite{shashkinPRL},
although much less pronounced, possibly because of the much higher
temperature.

\begin{figure}
\begin{center}
\includegraphics[width=0.7\columnwidth]{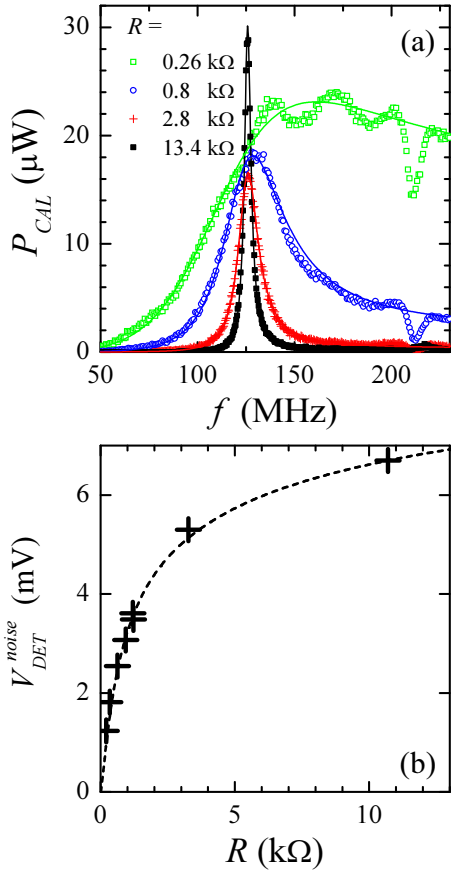}
\end{center}
\caption{Fig.~\ref{fig3}. (a) Frequency response spectra of the
setup at $I=0$ acquired with a spectrum analyzer for a set of
linear response resistances $R$ indicated in the legend.
Experimental data and fits are shown by symbols and lines,
respectively. (b) Calibration of the setup via equilibrium noise
measurement (see text). Symbols and line: measured chop amplitude
of the detector voltage and fit, respectively.}\label{fig3}
\end{figure}

The main ingredient of the shot noise measurements is the
calibration of the setup gain and bandwidth described below. A
rectified voltage $V_{\rm DET}$ at the output of the detector is
proportional to the power $P$ incident on the detector. The power
$P$ is the sum of contributions $P_T$ proportional to the
transistor noise $S_I^{T}$, $P_{Z_0}$ proportional to the input
current noise of the LTAmp $S_I^{Z_0}$ and a constant contribution
coming from the input voltage noise of the LTAmp and noises of all
other amplifiers. The LTAmp's input current noise is dominated by
the Johnson-Nyquist noise of the input resistor
$S_I^{Z_0}\approx4k_BT_{LTA}/Z_0$. The actual noise temperature
can be somewhat higher owing to extra current noise from the
active parts of the LTAmp. Small noises from the two resistors at
the input divider (fig.~\ref{fig1}) are neglected. Below we are
interested only in a differential part of $V_{\rm DET}$ measured
with a lock-in, which depends on the transistor S-D current $I$
and/or its (linear or differential) S-D resistance. Hence, up to
an unimportant constant one gets:
\begin{equation}
\begin{split}
V_{DET}^{noise} = D(P_T+P_{Z_0}) = S_I^{T}R^2\cdot DG\int |k_T|^2df+\\
 +S_I^{Z_0}Z_0^2\cdot DG\int |k_{Z_0}(f,R)|^2df, \label{eq1} \end{split}
\end{equation}
where $D$ is the detector power to voltage conversion coefficient,
$G$ -- total power gain of the amplifiers divided by the input
resistance $Z_0$. $k_T$ and $k_{Z_0}(f,R)$ denote the voltage
transfer functions of the corresponding noise sources to the input
of the LTAmp, which both depend on $f$ and $R$. The transfer
functions are set by the  parameters of the circuit in
fig.~\ref{fig1}, which we determine via the following calibration
procedure. We apply an rf signal of amplitude  $U_{CAL}$ at a
frequency $f$ to the input CAL IN (see fig.~\ref{fig1}) and
measure the contribution $V_{DET}^{CAL}$ to the output detector
voltage. The detector signal is proportional to power $P_{CAL}$
incident on the detector:
\begin{equation}
V_{DET}^{CAL}=DP_{CAL} =D|k^\prime_T|^2\cdot |k_{GEN}|^2G\cdot
U_{CAL}^2, \label{eq2}
\end{equation}
where $k_{GEN}$ is the voltage divider coefficient at the CAL IN
input of the circuit (fig.~\ref{fig1}). $k^\prime_T$ is the
rf-voltage  transfer function, which is related to the noise
transfer function $k_T$ from eq.~(\ref{eq1}) as
$|k^\prime_T|^2=\alpha |k_T|^2$. Factor
$\alpha=1+4\pi^2f^2R^2C_{SD}^2$ accounts for the suppression of
the transistor voltage noise caused by a stray S-D capacitance
$C_{SD}$. The power $P_{CAL}$ measured with a spectrum analyzer is
shown in fig.~\ref{fig3}a as a function of $f$ for a set of gate
voltages (symbols). The quality factor of the $LC$-resonator in
fig.~\ref{fig1} increases with $R$, which results in narrower peak
for more depleted channel in fig.~\ref{fig3}a. Solid lines
represent the best fits to the data used to accurately determine
the values of $L\approx 300$~nH, $C\approx1.5~$pF, the
drain-ground stray capacitance $C_D\approx3.9$~pF and
$C_{SD}\approx0.2~$pF. We find that $C_D$ is dominated by a stray
capacitance of the hand-made inductor ($L$ in fig.~\ref{fig1}) and
the value of $C_{SD}$ is close to an intrinsic parameter of the
transistor. The quality of the fits is almost perfect, apart from
small oscillations presumably caused by resonances in the
rf-tract. These discrepancies are not important as they occur
beyond the bandwidth used for noise measurements.

Fitting the data of fig.~\ref{fig3}a with eq.~(\ref{eq2}) returns
the value of the product $|k_{GEN}|^2G$, whereas a separate
knowledge of $G$ is required for noise measurements. This is
achieved via a measurement of the equilibrium noise of the setup,
which depends on $R$ (see eq.~(\ref{eq1})). We chop the transistor
gate voltage between the two values, corresponding to nearly zero
($\sim10\Omega$) and finite $R$ and measure the first harmonic ac
component of the detector voltage $V_{DET}^{noise}$ with a
lock-in. The result is plotted in fig.~\ref{fig3}b as a function
of $R$ (symbols). The noise signal $V_{DET}^{noise}$ increases as
the transistor is depleting, which reflects the increase of the JN
voltage noise of the transistor. The overall dependence is caused
by the interplay of the $R$-dependent bandwidth of the resonant
circuit and a (negative) contribution from the chopped LTAmp's
input current noise. The experimental behavior is well captured by
the dashed line fit in fig.~\ref{fig3}b. The fitting parameters
include $G$, the LTAmp's input noise temperature of
$T_{LTA}\approx4.7$K and the sample temperature of $T=4.2$K.
Consistently, an independent thermometry returned the value of
$\approx$5.3K for the temperature of the resistor $Z_0$
(fig.~\ref{fig1}). Under assumption of $T_{LTA}=5.3$K the best fit
to the data of fig~\ref{fig3}b would be obtained for a sample
temperature of $\approx4.8$K. This discrepancy represents a
possible systematic error in our calibration and shot noise
measurements, so that the Fano factor values given below may
actually be within 10\% higher.
\begin{figure}
\begin{center}
\includegraphics[width=0.7\columnwidth]{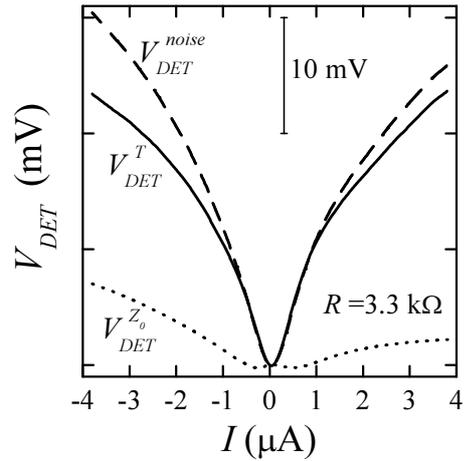}
\end{center}
\caption{Fig.~\ref{fig6}. Integrated detector voltage
$V_{DET}^{noise}=V_{DET}^T+V_{DET}^{Z_0}$ as a function of S-D
current (dashed line). Contributions from the transistor noise
($V_{DET}^T$) and input current noise of the LTAmp
($V_{DET}^{Z_0}$) are shown by solid and dotted lines,
respectively. The data are taken for the same gate voltage as the
traces shown by squares in fig~.\ref{fig2} (linear-response
resistance indicated in the figure)}\label{fig6}
\end{figure}

As follows from fig.~\ref{fig3}, at $I=0$ the rf-response of the
transistor and its equilibrium noise are successfully described by
a single stray capacitance parameter $C_{SD}$. We find that this
is not the case under non-linear transport conditions. Presumably,
the reason is the inhomogeneous electron density distribution
below the gate at $I\neq0$, which can change, e.g., distributed
gate-drain and gate-source capacitances. Instead of introducing
more fitting parameters at $I\neq0$, we calibrate the noise
transfer function $k_T$ in-situ. According to eq.~(\ref{eq2})
integration of the frequency response of the setup to the external
rf-signal gives:
\begin{equation}
DG\int|k_T|^2df=(\alpha |k_{GEN}|^2U_{CAL}^2)^{-1}\int
V_{DET}^{CAL}df \label{eq3}
\end{equation}
The quantity $K_{CAL}\equiv DG\int|k_T|^2df$ obtained in this way
accounts for the $I$-dependent gain and bandwidth of the shot
noise measurement in the nonlinear regime.


Using equations~(\ref{eq1}) and (\ref{eq3}) the measurement of the
shot noise spectral density $S_I^T$ is straightforward. In the
nonlinear regime the voltage noise of the transistor is determined
by the differential resistance $R_{diff}$ which substitutes $R$ in
eq.~(\ref{eq1}). The LTAmp's noise transfer function
$k_{Z_0}(f,R)$ is evaluated with the known $I=0$ circuit
parameters and $R=R_{diff}$. We measure the dc contribution to the
detector voltage caused by finite S-D current $V_{DET}^{noise}$.
This is achieved via a lock-in measurement of the derivative
$dV_{DET}/dI$ and subsequent numeric integration. The integration
constant is obtained via the equilibrium noise measurement
(fig.~\ref{fig3}b). The (arbitrarily offset) result is shown in
fig.~\ref{fig6} for one value of $V_g$ (see caption). Here, the
dashed line is the experimental $V_{DET}^{noise}$. The evaluated
contribution of the LTAmp's input current noise $V_{DET}^{Z_0}$ is
shown by dots. $V_{DET}^{Z_0}$ is not a constant, thanks to the
$R_{diff}$ dependence on $I$ in the nonlinear regime, which
slightly modifies the impedance connected to the input of the
LTAmp. As seen from fig.~\ref{fig2}, $R_{diff}$ changes stronger
for $I<0$, which results in a corresponding asymmetry of
$V_{DET}^{Z_0}$ as a function of $I$ in fig.~\ref{fig6}. The
difference $V_{DET}^T=V_{DET}^{noise}-V_{DET}^{Z_0}$ is the
contribution thanks to transistor shot noise shown by solid line.
The functional dependence of $V_{DET}^T$ on $I$ is related to that
of the noise spectral density as
$S_I^{T}=V_{DET}^T/(R_{diff}^2K_{CAL})$.

\begin{figure}
\begin{center}
\includegraphics[width=0.7\columnwidth]{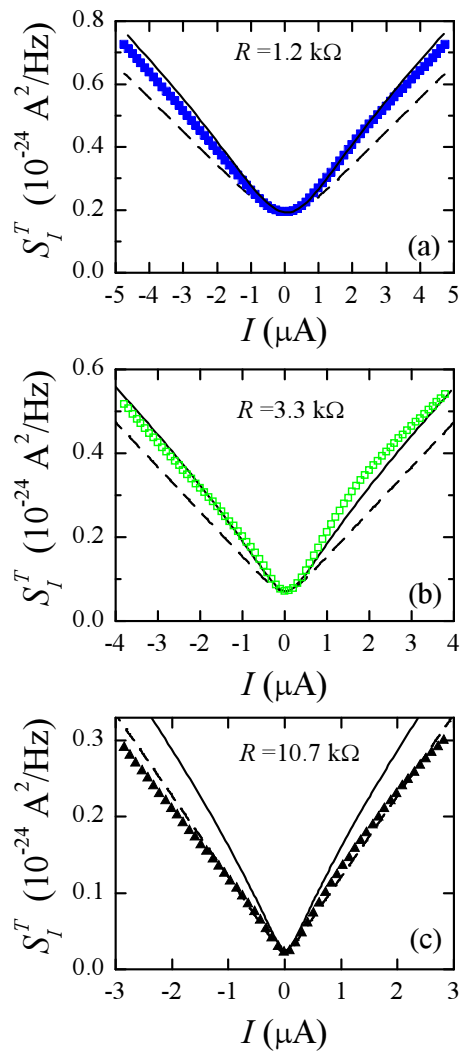}
\end{center}
\caption{Fig.~\ref{fig7}: Shot noise spectral density $S_I^T$ as a
function of S-D current for three values of the gate voltage
(linear-response resistance $R$ shown in the figures).
Experimental data in (a),(b),(c) are measured simultaneously with
the data of fig.~\ref{fig2} and are shown by the respective
symbols. Dashed lines are fits to the standard linear response
shot noise formula~\cite{henny}. Solid line are fits according to
eq.~(\ref{eq4}) and a model of nonlinear transport described in
the text.}\label{fig7}
\end{figure}

In figs.~\ref{fig7}a,\ref{fig7}b and \ref{fig7}c the noise
spectral density $S_I^{T}$ is plotted as a function of $I$ for
three values of the gate voltage (symbols). At $I=0$ the noise
spectral density is minimum and equals the JN value. At $I\neq0$,
$S_I^{T}$ increases as a function of $|I|$ and demonstrates a
nearly linear behavior at high enough currents. We find that for
all experimental traces, in the limit of high currents, the
differential Fano-factor is close to the universal value
$F_D\approx1/3$. For comparison, we plot $S_I$ expected for a
metallic diffusive conductor in the linear regime by dashed lines
in fig.~\ref{fig7}. These lines are drawn according to the
standard formula
$S_I=\frac{2}{3}R^{-1}\left[4k_BT+|eV|\coth(|eV|/2k_BT)\right]$,
where $R$ is the experimental linear response resistance and $V$
is the associated voltage drop $V=IR$~\cite{henny}. The symbols in
figs.~\ref{fig7}a and \ref{fig7}b (obtained for a less depleted
transistor) are systematically above the dashed lines, i.e. for
the same $I$ the noise spectral density exceeds the one obtained
with the above formula. Hence, at low currents the differential
Fano-factor is enhanced compared to 1/3, which is most pronounced
for the data in fig.~\ref{fig7}b with $F_D\approx0.5$ for
$|I|\sim1\mu$A. This discrepancy is beyond the experimental
uncertainty and can be explained by the effect of thermal
fluctuations in the nonlinear transport regime, as we propose
below.

The general result for shot noise spectral density of a
two-terminal conductor is usually express in terms of the
energy-dependent 1D eigen-channel transparencies
$T_n(E)$~\cite{blanterreview}. For the case of wide channel
transistor, it is convenient to express the same result in terms
of the energy-dependent conductance $\sigma(E)=e^2/h\sum T_n(E)$,
where $h$ is the Planck's constant, and the Fano-factor of the
partition noise ${F= \overline{T_n(1-T_n)}/\overline{T_n}}$:
\begin{equation}
\begin{split}
S_I=2\int \sigma(E)dE\{
f_L(1-f_L)+\\+f_R(1-f_R)+F(f_L-f_R)^2\}\label{eq4}
\end{split}
\end{equation}
Here, $f_i=(1+\exp[(E-\mu_i)/k_BT])^{-1}$ are the Fermi
distributions of the left ($i=L$) and right ($i=R$) reservoirs,
with respective electrochemical potentials of $\mu_i=\pm |e|V/2$.
The first two terms in the integrand of eq.~(\ref{eq4}) represent
the thermal noise of the reservoirs, while the last term stands
for the shot noise. Eq.~(\ref{eq4}) provides a phenomenological
description of the shot noise behavior in the nonlinear regime. We
find that the experimental enhancement of the differential
Fano-factor $F_D>1/3$ (fig.~\ref{fig7}a and~\ref{fig7}b) is not
necessarily related to energy dependence of $F$. Below we assume
that the partition noise Fano-factor in the last term of
eq.~(\ref{eq4}) has a universal value $F=1/3$ for metallic
diffusive conductors independent of energy.

The energy dependent conductance $\sigma(E)$ is directly related
to the transport current as ${I=|e|^{-1}\int\sigma(E)(f_L-f_R)dE}$
and can be obtained by fitting the experimental $I$-$V$ curves
(fig.~\ref{fig2}b). We model $\sigma(E)$ by a step-like functional
dependence on energy $\sigma(E)=\sigma_0(1+\tanh[(E-E_0-\lambda
|e|V)/\Delta])$. Parameters $\sigma_0\sim10^{-2}\Omega^{-1}$ and
$E_0,\Delta\sim1$~meV define the shape of the conductance step,
whereas the parameter $\lambda\sim0.1$ accounts for the asymmetry
of the $I$-$V$ curves (fig.~\ref{fig2}b) thanks to capacitive
effects of finite bias. This conductance model was chosen for its
analogy to the step-like behavior of the density of states near
the metal-insulator transition in two dimensions~\cite{ando}.
Fig.~\ref{fig2}b demonstrates that the model provides good fits
(dashed lines) to the experimental nonlinear $I$-$V$ curves
(symbols). The so-obtained $\sigma(E)$ and eq.~(\ref{eq4}) predict
the behavior of the noise spectral density, which is shown by
solid lines in fig.~\ref{fig7}.

The enhancement of the differential Fano-factor in a less depleted
channel at small $I$ (symbols in figs.~\ref{fig7}a
and~\ref{fig7}b) is qualitatively captured by the fits according
to eq.~(\ref{eq4}) (solid lines). Note, that thanks to assumption
of $F=1/3$ the eq.~(\ref{eq4}) reduces to $S_I=2/3|eI|$ at $T=0$.
In other words, within our model the enhancement of $F_D>F=1/3$ is
a finite temperature effect. The finite temperature determines the
thermal fluctuations in the electron flow incident on the
conductor. In the nonlinear regime, the contribution of thermal
fluctuations to the current noise increases as a function of $I$,
thanks to the energy dependence of the conductance $\sigma(E)$.
This results in enhanced differential Fano-factor $F_D>F=1/3$.
Note, that this nonlinear effect is not related to a
thermalization of the non-equilibrium carriers inside the
conductor~\cite{nagaev} or in the reservoirs~\cite{henny}.

The importance of the thermal fluctuations is best illustrated in
the ultimate case of thermally activated conductance in the
insulating phase, which in our model is achieved for
$E_0,\Delta\gg T$. This is expected to occur in a strongly
depleted transistor. Here, the eq.~(\ref{eq4}) predicts $F_D=1$ at
low $|I|$. In this case, the Poissonian value of the Fano-factor
is caused by a classical reason that the occupation number of the
current carrying states is small $f_i(E\sim E_0)\ll1$ ($i=L,R$).
This is fully analogous to the case of the shot noise for
thermionic emission in a vacuum tube considered by
Schottky~\cite{schottky}, and is not related to particular
properties of the model we used for $\sigma(E)$. In the limit of
high currents, the eq.~(\ref{eq4}) predicts a crossover to the
partition noise in the insulator breakdown regime with
$F_D\rightarrow F$. Unfortunately, we could not observe such a
behavior experimentally. Fig.~\ref{fig7}c shows the noise spectral
density for the lowest gate voltage (symbols), where the
non-linearities are most pronounced (same symbols in
fig.~\ref{fig2}). Unlike the prediction of eq.~(\ref{eq4}) (solid
line), $F_D\approx1/3$ and the data falls close to the standard
metallic linear-response result (dashed line). Note, however, that
at low $|I|$ this behavior might be an artefact caused by the
uncertainty in the voltage noise suppression factor $\alpha$,
which is most crucial for measurements at high
$R_{diff}\gtrsim3k\Omega$.

In summary, we performed the shot noise measurements in a
commercial high-frequency transistor near pinch-off. The
dependence of the shot noise on current is slightly nonlinear. The
differential Fano-factor is about $F_D\approx1/3$ in the limit of
high currents, and somewhat enhanced above the universal metallic
value at lower currents $1/3<F_D<0.5$. The model of nonlinear
transport near the pinch-off is suggested, which allows to explain
the results in terms of classical effect of thermal fluctuation,
without assuming the enhancement of the partition noise.

We acknowledge the discussions with A.A.~Shashkin, V.T.~Dolgopolov and V.F.~Gantmakher. Financial support by RFBR and the grant
MK-3470.2009.2 is gratefully acknowledged. VSK acknowledges
support from the Russian Science Support Foundation.

\end{document}